\documentstyle[preprint,aps]{revtex} 
\begin{document} 
\draft
\title{Nonlinear Modes of Liquid Drops as Solitary Waves} 
\author{A. Ludu and J. P. Draayer} 
\address{Department of Physics and Astronomy, Louisiana State University, 
Baton Rouge, LA 70803-4001} 
\date{\today} 
\maketitle 
\begin{abstract}
The nolinear hydrodynamic equations of the surface of a liquid drop
are shown to be directly connected to Korteweg de Vries (KdV, MKdV) systems,
giving traveling solutions that are cnoidal waves. They generate multiscale
patterns ranging from small harmonic oscillations (linearized model), to
nonlinear oscillations, up through solitary waves. These non-axis-symmetric
localized shapes are also described by a KdV Hamiltonian system. Recently such
``rotons'' were observed experimentally when the shape oscillations of a
droplet became nonlinear. The results apply to drop-like systems from cluster
formation to stellar models, including hyperdeformed nuclei and fission. 
\end{abstract} 
\narrowtext 
A fundamental understanding of non-linear oscillations of a liquid
drop (NLD), which reveals new phenomena and flows more complicated
than linear theory suggests, is needed in diverse areas of science and
technology.  Besides their direct use in rheological and surfactant
theory \cite{col,mit,th,elast,orig,sol,exp}, such models apply to cluster
physics \cite{nucl}, super- and hyper-deformed nuclei \cite{col}, 
nuclear break-up and fission \cite{mit,th,nucl}, 
thin films \cite{holt1}, radar \cite{elast} and even stellar masses and
supernova \cite{col,duri}. Theoretical approaches are
usually based on numerical calculations within different NLD models,
\cite{mit,th,elast} and explain/predict axis-symmetric, non-linear
oscillations that are in very good agreement with experiment
\cite{col,orig,sol,exp}. However, there are experimental results which show
non-axis-symmetric modes; for example, traveling rotational shapes
\cite{orig,sol} that can lead to fission, cluster emission, or fusion \cite
{orig,sol,exp}. 

In this letter the existence of analytic solutions of NLD models that
give rise to traveling solutions which are solitary waves is proven.
Higher order non-linear terms in the deviation of the shape from a
sphere produce surface oscillations that are cnoidal waves
\cite{book}. By increasing the amplitude of these oscillations, the
non-linear contribution grows and the drop's surface, under special
conditions (non-zero angular momentum), can transform from a cnodial
wave form into a solitary wave. This same evolution can occur if there
is a non-linear coupling between the normal modes. Thus this approach
leads to a unifying dynamical picture of such modes; specifically, the
cnoidal solution simulates harmonic oscillations developing into
anharmonic ones, and under special circumstances these cnoidal wave
forms develop into solitary waves. Of course, in the linear limit the
theory reproduces the normal modes of oscillation of a surface. 

Two approaches are used: Euler equations \cite{mit,th}, and
Hamiltonian equations, which describe the total energy of the system
\cite{mit}.  We investigate finite amplitude waves, for which the relative
amplitude is smaller than the angular half-width. These excitations are
also ``long'' waves, important in the cases of externally driven
systems, where the excited wavelength depends by the driving
frequency. The first original observations of travelling waves on liquid drops
are described in \cite{orig}. Similar travelling or running waves are also
discussed or quoted in \cite{mit,sol}.  These results suggest that higher
amplitude non-linear oscillations can lead to a traveling wave that
originates on the drop's surface and developes towards the interior. This
is shown to be related in a simply way to special solitary wave solutions,
called ``rotons'' in the present analysis. Recent experiments and numerical
tests\cite{nucl,la} suggest the existence of stable traveling waves for a
non-linear dynamics in a circular geometry, re-enforcing the theory. 

A new NLD model for describing an ideal, incompressible
fluid drop exercising irrotational flow with surface
tension, is employed in the analysis. Series expansion in terms of
spherical harmonics are replaced by localized, nonlinear shapes shown
to be analytic solutions of the system. The flow is potential and
therefore governed by Laplace's equation for the potential flow,
$\triangle \Phi =0$, while the dynamics is described by Euler's
equation,
\begin{equation}
\rho (\partial _{t}{\vec v}+({\vec v}\cdot \nabla  ){\vec v})
=-\nabla P+{\vec f}, \label{euler}
\end{equation}
where $P$ is pressure. If the density of the external force field is
also potential, ${\vec f} =-\nabla \Psi$ where $\Psi $ is proportional
to the potential (gravitational, electrostatic, etc.), then Eq.\
(\ref{euler}) reduces to Bernoulli's scalar equation. The boundary
conditions (BC) on the external free surface of the drop, $\Sigma 1$, and
on the inner surface $\Sigma 2$, \cite{mit,th,book},
are $\dot r | _{\Sigma 1}=(r_t + r_{\theta} \dot \theta 
+r_{\phi }\dot \phi ) | _{\Sigma 1}$ and $\dot r | _{\Sigma 2} =0$,
respectively. $\Phi _r =\dot r$ is the radial velocity, $\Phi _{\theta}=r^2
\dot{\theta}$, $\Phi _{\phi }=r^2 \sin \theta \dot {\phi}$ are the tangential
velocities.  The second BC occurs only in the case of fluid shells or bubbles.
A convenient geometry places the origin at the center-of-mass of the
distributon $r(\theta , \phi , t)=R_0 [1+g(\theta )\eta (\phi -Vt)]$ 
and introduces for the dimensionless shape function $g\eta$
a variable denoted $\xi $.
Here $R_0$ is the radius of the undeformed spherical drop and $V$ is the
tangential velocity of the traveling solution $\xi $ moving in the
$\phi $ direction and having a constant transversal profile $g$ in the
$\theta $ direction. The linearized form of the first BC, 
$\dot r | _{\Sigma 1}={r_{t}} | _{\Sigma 1}$, allows only
radial vibrations and no tangential motion of the fluid on $\Sigma 1$,
\cite{mit,th,book}. The second BC restricts the radial flow
to a spherical layer of depth $h(\theta )$ by requiring $\Phi _r
|_{r=R_0 -h}=0$. This condition stratifies the flow in the surface
layer, $R_0 -h \leq r \leq R_0 (1+\xi )$, and the liquid bulk $r\leq
R_0 -h$. In what follows the flow in the bulk will be considered
negligible compared to the flow in the surface layer. This condition
does not restrict the generality of the argument because $h$ can
always be taken to be $R_0$. Nonetheless, keeping $h < R_0$
opens possibilities for the investigation of more complex fluids, e.g. 
superfluids, flow over a rigid core, multilayer systems 
\cite{mit,exp} or
multiphases, etc.  Instead of an expansion of $\Phi$ in term of
spherical harmonics, consider the following form
\begin{equation} 
\Phi=\sum_{n=0}^{\infty}
(r/R_{0}-1)^{n}f_{n}(\theta , \phi ,t). \label{series} 
\end{equation} 
The convergence of the series is
controlled by the value of the small quantity $\epsilon =max|{{r-R_0 }
\over {R_0 }}|$, \cite{book}. The condition $max |h/R_0 |\simeq
\epsilon $ is also assumed to hold in the following development. 
Laplace's equation introduces a system of recursion relations for the
functions $f_n $, $f_2 =-f_1 -\triangle _{\Omega }f_0 /2$, etc.,
where $\triangle _{\Omega }$ is the $(\theta ,\phi )$ part of the
Laplacean. Hence the set of unknown $f_n$'s reduces to 
$f_0 $ and $f_1$.  The second BC, plus
the condition $\xi _{\phi } = -V\xi _{t}$,
for traveling waves, yields to second order in $\epsilon$,
\begin{equation}
f_{0,\phi }=VR_{0}^{3}\sin ^{2}\theta \xi(1+2\xi )/h
+{\cal O}_{3}(\xi ), \label{final}
\end{equation}
i.e., a connection between the flow potential and the shape, 
which is typical of nonlinear systems. 
Eq.(3) together with the relations $f_1 \simeq R_{0}^{2}\xi_{t}\simeq
{{2h}\over {R_0 }}f_2 \simeq -{{h\triangle _{\Omega }f_0 }\over
{R_{0}+2h}}$, which follow from the BC and recursion, characterize the flow as
a function of the surface geometry. The balance of the dynamic and
capillary pressure across the surface $\Sigma 1$ follows by expanding
up to third order in $\xi $ the square root of the surface energy of
the drop \cite{mit,th,book},
\begin{equation}
U_S =\sigma R_{0}^{2}\int _{\Sigma 1}
(1+\xi )\sqrt{(1+\xi )^2 +\xi _{\theta }^{2}
+\xi _{\phi }^{2}/\sin ^{2} \theta }d\Sigma , \label{surfenerg}  
\end{equation}
and by equating its first variation with the local mean curvature of
$\Sigma 1$ under the restriction of the volume conservation. The
surface pressure, in third order, reads
\begin{equation}
P|_{\Sigma 1}={{\sigma } \over {R_0 }}
(-2\xi-4\xi ^2 -\triangle _{\Omega }\xi +3\xi \xi ^{2}_{\theta }
ctg \theta ), \label{press} 
\end{equation}
where $\sigma$ is the surface pressure coefficient and the terms $\xi
_{\phi ,\theta}, \xi_{\phi,\phi }$ and $\xi _{\theta ,\theta }$ are
neglected because the relative amplitude of the deformation $\epsilon
$ is smaller than the angular half-width $L$, $\xi= \xi _{\phi \phi
}\simeq \epsilon ^{2}/L^2 \ll 1$, as most of the experiments
\cite{sol,exp,holt1,la} concerning traveling surface patterns show. 
Eq.(5) plus the BC
yield, to second order in $\epsilon$,
\begin{eqnarray}
\Phi _{t}|_{\Sigma 1}&+&{{V^2 R_{0}^{4}\sin ^2 \theta }\over {2h^2 }} \xi ^2
\nonumber \\  &=& {{\sigma }\over {\rho R_0 }}(2\xi + 4\xi ^2+\triangle
_{\Omega }
\xi -3\xi ^2 \xi _{\theta }ctg \theta ). \label{dyn1}
\end{eqnarray} 
The linearized version of
Eq.\ (\ref{dyn1}) together with the linearized BC, 
$\Phi _{r}|_{\Sigma 1}=R_0 \xi _{t}$, yield a
limiting case of the model, namely, the normal modes of oscillation of
a liquid drop with spherical harmonic solutions
\cite{mit,th}.  Differentiation of Eq.\ (\ref{dyn1}) with
respect to $\phi $ together with Eqs.(3,5) yields the dynamical
equation for the evolution of the shape function $\eta (\phi -Vt)$: 
\begin{equation}
A\eta _t +B\eta _{\phi }+Cg\eta \eta _{\phi }+D\eta _{\phi \phi \phi
}=0, \label{kdv}
\end{equation}
which is the Korteweg-de Vries (KdV) equation \cite{book} with
coefficients depending parametrically on $\theta $
$$
A=V{{R_{0}^{2}(R_0 +2h)\sin ^2 \theta }\over {h}}, \ \
B=-{{\sigma }\over {\rho R_0 }}{{(2g+\triangle _{\Omega }g)} \over {g}},
$$
\begin{equation}
C=8\biggl ( {{V^2R_{0}^{4}\sin ^4 \theta } \over {8h^2 }}-{{\sigma }
\over {\rho R_{0}}} \biggr ) , \ \ \ D=-{{\sigma }\over {\rho
R_{0}\sin ^2 \theta }} . \label{coef}
\end{equation}
In the case of a two-dimensional liquid drop, the
coefficients in Eq.\ (\ref{coef}) are all constant. Eq.\ (\ref{kdv})
has traveling wave solutions in the $\phi $ direction if $Cg/(B-AV)$
and $D/(B-AV)$ do not depend on $\theta $. These two conditions
introduce two differential equations for $g(\theta )$ and $h(\theta )$
which can be solved with the boundary conditions $g=h=0$ for $\theta
=0, \pi $. For example, $h_1 =R_0 \sin ^2 \theta $ and $g_1 =
P_{2}^{2} (\theta )$ is a particular solution which is valid for $h
\ll R_0 $. It represents a soliton with a quadrupole transvere 
profile, being in good agreement with \cite{mit,sol}.  The
next higher order term in Eq.\ (\ref{dyn1}), $-3\xi ^2 \xi _{\theta }
ctg \theta $, introduces a $\eta ^2 \eta _{\phi }$ nonlinear term into
the dynamics and transforms the KdV equation into the modified KdV
equation \cite{book}. The traveling wave solutions of Eq.\ (\ref{kdv}) are
then described by the Jacobi elliptic function \cite{book}
\begin{equation}
\eta ={\alpha }_{3}+({\alpha }_{2}-{\alpha }_{3})sn^2 \biggl (
\sqrt{{{C(\alpha _3 -\alpha _2 )} \over {12D}} }(\phi -Vt) ; m\biggr )
, \label{cnoidal}
\end{equation}
where the $\alpha _i$ are the constants of integration introduced
through Eq.\ (\ref{kdv}) and are related through the velocity $V=C(\alpha
_1 +\alpha _2 +\alpha _3 )/3A+B/A$ and $m^2 ={{\alpha _3 -\alpha
_2}\over {\alpha _3 - \alpha _1}}$. $m \in [0,1]$ is the free parameter of
the elliptic $sn$ function. This result for Eq.\ (\ref{cnoidal}) is
known as a cnoidal wave solution with angular period
$T=K[m]\sqrt{C(\alpha _3 -\alpha _1 )/3D}$ where $K(m)$ is the Jacobi
elliptic integral. If $\alpha _2 \rightarrow \alpha _1 \rightarrow 0$, then
$m\rightarrow 1, T \rightarrow \infty$ and a one-parameter ($\eta
_0$) family of traveling pulses (solitons or anti-solitons) is
obtained,
\begin{equation}
\eta _{sol}=\eta _0 sech ^2 [(\phi -Vt)/L], \label{soli}
\end{equation}
with velocity $V=\eta _0 C/3A+B/A$ and angular half-width
$L=\sqrt{12D/C\eta _0}$. Taking for the coefficients $A$ to $D$
the values given in Eq.(8) for $\theta =\pi /2$ (the equatorial cross 
section) and $h_1$, $g_1$ from above, one can calculate numerical values 
of the parameters of any roton excitation function of $\eta _0$ only.

The soliton, among other wave patterns, has a special shape-kinematic 
dependence
$\eta _0 \simeq V\simeq 1/L $; a higher soliton is narrower and travels
faster. This relation can be used to
experimentally distinguish solitons from other modes or turbulence. 
When a layer thins ($h\rightarrow 0$) the coefficient $C$ in eq.(8) 
approaches zero on average, producing a break in the
traveling wave solution ($L$ becomes singular) because of the 
change of sign under the square 
root, eq.(9). Such wave turbulence from capillary waves on thin 
shells was
first observed in [9]. For the water shells described there, eq.(8) gives
$h(\mu m)\leq 20\nu /k$, that is $h$=15-25$\mu$m at 
$V$=2.1-2.5m$s^{-1}$
for the onset of wave turbulence, in good agreement with the abrupt
transition experimentaly noticed. The cnoidal solutions provide the
nonlinear wave interaction and the transition from competing linear wave
modes ($C\le 0$) to turbulence ($C\simeq 0$). In the KdV eq.(7), the
nonlinear interaction balances or even dominates the linear damping and 
the cnoidal (roton) mode occurs as a bend mode ($h$ small and coherent
traveling profile) in agreement with [9]. 
The condition for the existence of a positive amplitude soliton is $gCD\geq 0$
which, for $g\leq 0$, limits the velocity from below to the value
$V \geq  h \omega _2 /R_{0}$
where $\omega _2$ is the Lamb frequency 
for the $\lambda =2$ linear mode \cite{mit,th}. 
This inequality can be related to the ``independent running wave'' described
in  \cite{sol}, which lies close to the $\lambda =2$ mode.  Moreover,
since the angular group velocity of the $(\lambda ,\mu )$ normal mode,
$V_{\lambda ,\mu }=\omega _{\lambda }/\mu $, has practically the same
value for $\lambda =2$ ($\mu =0,\pm 1$, tesseral harmonics) and for
$\lambda =\mu $, any $\lambda $ (sectorial harmonics) this inequality 
 seems to be essential for any combination of rank 2 tesseral
or sectorial harmonics, in good agreement with the conclusions in
\cite{mit}. The periodic limit of the cnoidal wave is reached for $m\simeq
0$, that is, $\alpha _2 -\alpha _3 \simeq 0$, and the shape is
characterized by harmonic oscillations ($sn \rightarrow \sin$ in Eq.\
(\ref{cnoidal})) which realize the quadrupole mode of a linear theory
$Y_{2}^{\mu }$ limit \cite{mit,th} or the oscillations of tesseral
harmonics \cite{mit}, Fig. 1. 

The NLD model introduced in this paper yields a smooth transition from
linear oscillations to solitary traveling solutions (``rotons'') as a
function of the parameters $\alpha _{i}$; namely, a transition from
periodic to non-periodic shape oscillations.  In between these limits the
surface is described by nonlinear cnoidal waves. In Fig.1 the transition
from a periodic limit to a solitary wave is shown, in comparison with the
corresponding normal modes which can initiate such cnoidal nonlinear
behavior. This situation is similar to the transformation of the flow
field from periodic modes at small amplitude to traveling waves at larger
amplitude \cite{mit,sol}.  The solution goes into a final form if the
volume conservation restriction is enforced: $\int _{\Sigma }(1+g(\theta
)\eta (\phi ,t))^3 d\Omega =4 \pi $ and requires $\eta (\phi ,t)$
to be periodic. The periodicity condition, $nK[ (\alpha
_3 -\alpha _2 )/(\alpha _3 -\alpha _1 )] = \pi \sqrt {\alpha _3 -\alpha _1
}$ for any positive integer $n$, is only fulfilled for a finite number of
$n$ values, and hence a finite number of coresponding cnoidal modes. 
In the roton limit the
periodicity condition becomes a quasi-periodic one because the amplitude
decays rapidly.
This approach could be extended to describe elastic modes of
surface as well as their nonlinear coupling to capillary waves. The
double-periodic structure of the elliptic solutions \cite{book} could
describe the new family of normal wave modes predicted in
\cite{elast}. 

The development up to this point was based on Euler's equation. The same
result will now be shown to emerge from a Hamiltonian analysis of the NLD
system. Recently, Natarajan and Brown \cite{mit} showed that the NLD is a
Lagrangian system with the volume conservation condition being a Lagrange
multiplier. In the third order deviation from spherical, the NLD becomes a
KdV infinite-dimensional Hamiltonian system described by a nonlinear
Hamiltonian function $H=\int _{0}^{2\pi }{\cal H} d\phi $. In the linear
approximation, the NLD is a linear wave Hamiltonian system
\cite{mit,th}. If terms depending on $\theta $ are absorbed into
definite integrals (becoming parameters) the total energy is a function of
$\eta $ only.  Taking the kinetic energy from \cite{mit,th}, $\Phi $ from
Eq.\ (\ref{series}) and using the BC, the dependence of
the kinetic energy on the tangential velocity along $\theta $ direction,
$\Phi _{\theta }$, becomes negligible and the kinetic energy can be
expressed as a $T[\eta]$ functional. For traveling wave solutions
$\partial _t = -V \partial _{\phi }$, to third order in $\epsilon $, after
a tedious but feasible calculus, the total energy is: 
\begin{equation}
E=\int_{0}^{2\pi }({\cal C}_{1}\eta +{\cal C}_{2}{\eta }^{2}+{\cal
C}_{3}\eta ^{3} +{\cal C}_{4}{\eta }_{\phi }^{2} )d\phi , \label{e}
\end{equation} 
where 
${\cal C}_1 =2\sigma R_{0}^{2}S_{1,0}^{1,0}$, 
${\cal C}_{2}=\sigma R_{0}^{2}(S_{1,0}^{1,0}+S_{0,1}^{1,0}/2) + 
R_{0}^{6}\rho V^2
C_{2,-1}^{3,-1}/2$, ${\cal C}_{3}=\sigma R_{0}^{2}S_{1,2}^{1,0}/2+
R_{0}^{6}\rho V^2 (2S_{-1,2}^{3,-1}R_0+S_{-2,3}^{5,-2}+ R_0 
S_{-2,3}^{6,-2})/2$,
${\cal C}_{4}=\sigma R_{0}^{2}S_{2,0}^{-1,0}/2$, with
$S_{i,j}^{k,l}=R_{0}^{-l}\int_{0}^{\pi}h^{l}g^{i}g_{\theta }^{j}{\sin 
}^{k}\theta d\theta $.
Terms proportional to $\eta \eta _{\phi }^{2}$ can be neglected
since they introduce a factor $\eta _{0}^{3}/L^2 $ which is small compared
to $\eta _{0}^{3}$, i.e. it is in the third order in $\epsilon $.  If Eq.\
(\ref{e}) is taken to be a Hamiltonian, $E \rightarrow H[\eta ]$, then the
Hamilton equation for the dynamical variable $\eta $, taking the usual form
of the Poisson bracket, gives
\begin{equation}
\int_{0}^{2\pi }\eta _t d\phi =\int_{0}^{2\pi } 
(2{\cal C}_{2}\eta_ {\phi }+6{\cal C}_{3}\eta \eta
_{\phi }-2{\cal C}_{4} \eta _{\phi \phi \phi })d\phi. \label{eqh2}
\end{equation}
Since for the function $\eta (\phi -Vt)$
the LHS of Eq.(12) is zero, the integrand in the RHS gives the KdV equation.
Hence, the energy of the NLD model,
in the third order, is interpreted as a Hamiltonian of the 
KdV equation \cite{exp,book}. This is in full agreement with
the result finalized by Eq.\ (\ref{kdv}) for an appropriate choice 
of the parameters and the Cauchy conditions for $g,h$. 
The dependence of $E(\alpha _1 ,\alpha _2
)|_{Vol=constant}$, Eq.(11), shows an energy minimum in which the 
solitary waves are stable, \cite{la}.

The nonlinear coupling of modes in the cnoidal solution could explain 
the occurence of many resonances for the $l=2$ 
mode of rotating liquid drops, at a given (higher) 
angular velocity, \cite{shift}.
The rotating quadrupole shape is close to the soliton limit of 
the cnoidal wave.
On one hand, the existence of many resonances is a consequence of
by the multi-valley profile of the effective potential energy for the KdV, 
(MKdV) equation: $\eta _{x}^2=a\eta+b\eta ^2+c\eta ^3 +(d \eta ^4)$, 
\cite{book}. 
The frequency shift predicted by Busse in \cite{shift} can be reproduced in 
the present theory 
by choosing the solution $h_1 =R_0 sin \theta /2$. It results the same 
additional pressure drop in the form of $V^2 \rho R_{0}^{2}sin^2 \theta /2$
like in \cite{shift}, and hence a similar result. For a roton emerged from a 
$l=2$
mode, by calculating the half-width ($L_2$) and  amplitude ($\eta _{max,2}$)
which fitt the quadrupole shape it results a law for
the frequency shift:
$\Delta \omega _2 / \omega _2 =(1\pm 4L^2 (\alpha _3 
-\alpha _2 )/3R_0)^{-1}V/\omega _2$, showing a good agreement 
with the observations of Annamalai {\it {et al}} in \cite{shift}, i.e.
many resonances and nonlinear dependence of the shift on $\Omega =V$.
The special damping of the $l=2$ mode for rotating drops could also
be a consequence of the existence of the cnoidal solution. An increasing 
in the velocity $V$ produces a modification of the balance of the 
coefficients $C/D$ which is equivalent with an increasing in dispersion.

The model introduced in this article proves that traveling analytic
solutions exist as cnoidal waves on the surface of a liquid drop. These
traveling deformations (``rotons") can range from small oscillations
(normal modes), to cnoidal oscillations, and on out to solitary waves. 
The same approach can be applied to bubbles as well, except that the 
boundary condition on $\Sigma _2$ is replaced by a far-field condition [2,3]
(recently important in the context of single bubble sonoluminiscence).
Nonlinear phenomena can not be fully investigated with normal linear
tools, e.g. spherical harmonics. Using analytic non-linear solutions
sacrifices the linearity of the space but replaces it with multiscale
dynamical behavior, typical for non-linear systems (solitons, wavelets,
compactons \cite{la}).  
They can be applied to phenomena like cluster formation in
nuclei, fragmentation or cold fission, the dynamics of the pellet surface
in inertial fusion, stellar models, and so forth. 

\vskip .25cm
Supported by the U.S. National Science Foundation through a regular grant,
No. 9603006, and a Cooperative Agreement, No. EPS-9550481, that includes
matching from the Louisiana Board of Regents Support Fund.

\begin{figure} 
\caption{The cnoidal solution for $\theta = 0$. The soliton limit
and a 3- and 4-mode solution is shown. The closest spherical
harmonics to each of the cnoidal wave profiles (labelled Cn and Sol,
respectively) is given for comparison. The labels  
$\lambda ,\mu $ and the parameters $\alpha _{1,2,3}$
of the coresponding cnoidal solution are given.}
\label{fig1} 
\end{figure}

\end{document}